%% file: CIKM2021.tex
\PassOptionsToPackage{table,xcdraw}{xcolor}

\documentclass[sigconf,authorversion]{acmart}

\usepackage[table,xcdraw]{xcolor}
\usepackage{textcomp}
\usepackage{booktabs} % For formal tables
\usepackage{multicol}
\usepackage{enumitem}
\usepackage{enumerate}
\usepackage{multirow}
\usepackage{pifont}
\newcommand{\cmark}{\checkmark}%
\newcommand{\xmark}{$\times$}%
\usepackage[normalem]{ulem}
\useunder{\uline}{\ul}{}
\usepackage{subfigure}
\usepackage{balance}

\copyrightyear{2021} 
\acmYear{2021} 
\setcopyright{acmlicensed}\acmConference[CIKM '21]{Proceedings of the 30th ACM International Conference on Information and Knowledge Management}{November 1--5, 2021}{Virtual Event, QLD, Australia}
\acmBooktitle{Proceedings of the 30th ACM International Conference on Information and Knowledge Management (CIKM '21), November 1--5, 2021, Virtual Event, QLD, Australia}
\acmPrice{15.00}
\acmDOI{10.1145/3459637.3482486}
\acmISBN{978-1-4503-8446-9/21/11}

\usepackage{etoolbox}
\makeatletter
\apptocmd{\@mkauthors@iii}{%
  \global\setbox\mktitle@bx=\vbox{\noindent
    \box\mktitle@bx\par\vskip2pt\nointerlineskip
    {\centering\large
     FuxiCTR Leaderboard: \textcolor{blue!70!black}{\url{https://recbench.github.io/FuxiCTR}}\par}%
    \vskip16pt}%
}{}{\ClassWarning{CIKM2021}{Failed to patch \@mkauthors@iii}}
\makeatother

\begin{document}
\fancyhead{} 

%\title[EASE: Learning Lightweight Semantic Feature Adapters from Large Language Models for CTR Prediction]{EASE: Learning Lightweight Semantic Feature Adapters \\from Large Language Models for CTR Prediction}

\title{Open Benchmarking for Click-Through Rate Prediction}

% \subtitle{FuxiCTR Leaderboard: \textcolor[HTML]{2E639E}{\url{https://recbench.github.io/FuxiCTR}}}  % 已由导言区补丁改为在作者列表下方显示

\author{Jieming Zhu}
\authornote{Both authors contributed equally to this work.}
\affiliation{%
  \institution{Huawei Noah's Ark Lab}
  \city{Shenzhen}
  \country{China}
}
\affiliation{\texttt{jiemingzhu@ieee.org}}

\author{Jinyang Liu}
\authornotemark[1]
\authornote{Part of the work was done when the author studied at Sun Yat-Sen University.}
\affiliation{%
  \institution{The Chinese University of Hong Kong}
 \city{Hong Kong}
  \country{China}
}
\affiliation{\texttt{jyliu@cse.cuhk.edu.hk}}

\author{Shuai Yang}
\affiliation{%
\institution{Peking University}
  \city{Beijing}
  \country{China}
}
\affiliation{\texttt{ethanyang@pku.edu.cn}}

\author{Qi Zhang}
\affiliation{%
  \institution{Huawei Noah's Ark Lab}
  \city{Beijing}
  \country{China}
}
\affiliation{\texttt{zhangqi193@huawei.com}}

\author{Xiuqiang He}
\affiliation{%
  \institution{Huawei Noah's Ark Lab}
  \city{Shenzhen}
  \country{China}
}
\affiliation{\texttt{hexiuqiang1@huawei.com}\vspace{1.5ex}}

% https://www.overleaf.com/8426897472wkzxmswmkjsp

\begin{abstract}
%In many applications, such as recommender systems, online advertising, and product search,

Click-through rate (CTR) prediction is a critical task for many applications, as its accuracy has a direct impact on user experience and platform revenue. In recent years, CTR prediction has been widely studied in both academia and industry, resulting in a wide variety of CTR prediction models. Unfortunately, there is still a lack of standardized benchmarks and uniform evaluation protocols for CTR prediction research. This leads to non-reproducible or even inconsistent experimental results among existing studies, which largely limits the practical value and potential impact of their research. In this work, we build an open benchmark for CTR prediction, namely FuxiCTR-Bench, and present a rigorous comparison of different models in a reproducible manner. To this end, we ran over 7,000 experiments for more than 12,000 GPU hours in total to re-evaluate 24 existing models on multiple datasets and settings. Surprisingly, our experiments show that with sufficient hyper-parameter search and model tuning, many deep models have smaller differences than expected. The results also reveal that making real progress on the modeling of CTR prediction is indeed a very challenging research task. We believe that our benchmarking work could not only allow researchers to gauge the effectiveness of new models conveniently but also make them fairly compare with the state of the arts. We have publicly released the source code, evaluation protocols, and hyper-parameter settings of our work to promote reproducible research in this field.
\end{abstract}

%and sometimes are even inconsistent with what reported in the literature. 

%
% The code below should be generated by the tool at
% http://dl.acm.org/ccs.cfm



\keywords{Recommender Systems; Sponsored Search; Computational Advertising; CTR Prediction; Benchmarking}

\maketitle

\input{sections/1_introduction.tex}

\input{sections/2_background.tex}

\input{sections/4_experiment.tex}

\input{sections/5_relatedwork.tex}

\input{sections/6_conclusion.tex}

\balance
\bibliographystyle{ACM-Reference-Format}
\bibliography{CIKM2021}
\end{document}

%% file: sections/1_introduction.tex
\section{Introduction}
In many applications such as recommender systems, online advertising, and product search, click-through rate (CTR) is a key factor in business valuation. CTR prediction, which aims to accurately predict the probability of a user clicking or interacting with a candidate item, is an important task in industry. For applications with a large user base, even a small improvement in prediction accuracy can potentially lead to a large increase in the overall revenue. For example, existing studies from Google~\cite{DCN,WideDeep} and Microsoft~\cite{EnsembleBing} reveal that an absolute improvement of 1\textperthousand ~in logloss (or AUC) is considered as practically significant in real CTR prediction problems. 
In contrast to other data types such as images and texts, CTR prediction problems usually involve data of large scale and high sparsity and comprise many categorical features of different fields (e.g., billions of samples with millions of features in app recommendation of Google Play~\cite{WideDeep}). Therefore, it is a great challenge to make significant accuracy improvements in CTR prediction. 

%accurate predictions can facilitate better item ranking and thus maximize the revenue from user clicks and interactions.

The importance and unique challenges of CTR prediction have attracted a lot of research attention from both academia and industry. CTR prediction models have evolved from simple logistic regression (LR)~\cite{LR,FTRL}, factorization machines (FM)~\cite{FM,FFM} and decision trees~\cite{GBDTLR,XGBoost}, to deep neural networks (DNN)~\cite{DNN,WideDeep}. Notably, numerous deep models have been proposed and shown remarkable performance gains in industrial CTR prediction problems, such as Wide\&Deep~\cite{WideDeep}, DeepFM~\cite{DeepFM}, DCN~\cite{DCN}, xDeepFM~\cite{xDeepFM}, FiBiNET~\cite{FiBiNET}, DIN~\cite{DIN}, and so on.

Despite the success of these studies, there is still a lack of standardized benchmarks and uniform evaluation protocols for CTR prediction tasks. As a consequence, even though some common datasets (e.g., Criteo~\cite{Criteo} and Avazu~\cite{Avazu}) are used, existing studies often perform their own data partitions (e.g., with unknown train-test splitting or using unknown random seeds) and preprocessing steps (regarding how to cope with numeric features and how to filter rare categorical features). This leads to the non-reproducible and even inconsistent experimental results among these studies, since their nonstandard data preprocessing makes the results of any two different papers not comparable. Every work claims to achieve the best results with a significant improvement on their own data partition, yet no one knows what it would be like if the same evaluation protocol had been applied for fair comparisons. Due to the lack of open benchmarking results for reference, readers may doubt whether the baseline models in a paper are implemented correctly and tuned with rigour, but none of these studies report the details or open the source code of their baseline implementations. In some cases, the official or third-party source code of some popular models is available (e.g., DeepCTR library~\cite{DeepCTR}), but we found that the training details about hyper-parameter settings, data loading, and early stopping are usually missing, which makes it difficult to reuse the code for reproducing the existing results. Such non-reproducibility and inconsistency issues largely limit the practical value and potential impact of the research in this field. Moreover, due to the lack of reusable and comparable benchmarking results in the literature, researchers need to re-implement all the baselines and re-evaluate them on their own data partition when publishing a new paper. This is a tedious yet redundant effort, heavily increasing the burden of researchers to develop new models.

%Although each single paper may have a uniform experimental setting, inconsistent results may occur across different papers.

Inspired by the success of ImageNet benchmark~\cite{ImageNet} in the CV domain and GLUE benchmark~\cite{GLUE} in the NLP domain, in this paper, we propose to perform open benchmarking for CTR prediction. Our work not only standardizes the open benchmarking pipeline for CTR prediction, but also makes a rigorous comparison of different models for reproducible research. To this end, we ran over 7,000 experiments for more than 12,000 GPU hours with a uniform setup to re-evaluate 24 existing models on multiple dataset settings of the two widely-used datasets, including Criteo~\cite{Criteo} and Avazu~\cite{Avazu}. Our experiments show somewhat surprising results. After sufficient hyper-parameter search and model tuning, many recent models have smaller differences than expected and sometimes are even inconsistent with what was reported in the literature. A similar study~\cite{Recsys_worryanalysis} also performed a re-evaluation of multiple representative papers on recommender systems, raising reproducibility issues of the results and concerns about lacking sufficient optimization of the baselines used for comparison. In contrast to this work, we take one step further to build an open benchmark for CTR prediction, namely FuxiCTR-Bench, which serves as an essential benchmark of our RecBench project\footnote{RecBench Project: \textcolor{blue!70!black}{\url{https://recbench.github.io}}}. The goal is to set up a standardized open benchmarking pipeline for recommender systems research and to drive reproducible research in this field through opening the most comprehensive benchmarking results along with the well-documented reproducing steps. We believe that such a benchmarking study could be beneficial to multiple different groups of readers.
\begin{itemize}
    \item \textbf{Researchers}: The benchmark could not only help researchers analyze strengths and bottlenecks of existing models, but also allow them to gauge the effectiveness of new models conveniently. Moreover, our benchmark demonstrates some good practices to fairly compare with the state of the arts for future research.
    \item \textbf{Practitioners}: The availability of benchmarking code and results can help industrial practitioners assess the applicability of new research models in their own problems and allow them to try new models with little efforts on their own datasets. 
    \item \textbf{Competitors}: Leveraging our source code and hyper-parameters, competitors can easily implement high-performance baselines and ensembles in related competitions.
    \item \textbf{Beginners}: For beginners in this field, especially for students, our benchmarking code and the detailed reproducing steps can serve as a guidebook to learn the model implementations and model tuning tricks for CTR prediction. It is also valuable to apply our project 
    for educational purposes.
\end{itemize}
    
In summary, our work makes the following main contributions:
\begin{itemize}
    \item To the best of our knowledge, our work makes the first step towards open benchmarking for CTR prediction.
    \item We open all the benchmarking code, evaluation protocols, hyper-parameter settings, and experimental results on the website to foster reproducible research on CTR prediction.
    \item Our work reveals the non-reproducibility and inconsistency issues in existing studies, and calls for openness and rigour of model evaluation in future research on CTR prediction.
\end{itemize}

The remainder of this paper is organized as follows. Section~\ref{sec_background} introduces the overview of CTR prediction. Section~\ref{sec:exp} describes our evaluation protocols and benchmarking results. Further, some discussions are made in Section~\ref{sec:discussion} and the related work is reviewed in Section~\ref{sec:relatedwork}. Finally, we conclude the paper in Section~\ref{sec:conclusion}.

%% file: sections/2_background.tex
\section{CTR Prediction}\label{sec_background}
In this section, we provide an overview of CTR prediction and then briefly review some of the representative models.

\subsection{Overview}\label{sec:ctr}
The objective of CTR prediction is to predict the probability that a user will click a given item. How to improve the accuracy of CTR prediction remains a challenging research problem. In contrast to other data types, such as images and texts, data in CTR prediction problems are typically in tabular format, comprising either numerical, categorical, or multi-valued (or sequence) features of multiple different fields. The sample size is often large, yet the feature space is highly sparse. For example, app recommendation in Google Play~\cite{WideDeep} involves billions of samples with millions of features. 
In general, a CTR prediction model consists of the following key parts.

\subsubsection{Feature Embedding}
Input instances for CTR prediction generally contain three groups of features, \textit{i.e.}, user profile, item profile, and context information. Each group has a number of fields as follows:
\begin{itemize}
    \item \textit{User profile}: age, gender, city, occupation, interests, etc. 
    \item \textit{Item profile}: item ID, category, tags, brand, seller, price, etc. 
    \item \textit{Context}: weekday, hour, position, slot id, etc. 
\end{itemize}
Features in each field may be categorical, numeric, or multi-valued (e.g., multiple tags of a single item). Since most features are very sparse, leading to high-dimensional feature space after one-hot or multi-hot encoding, it is common to apply feature embedding to mapping these features into low-dimensional dense vectors. We summarize the embedding process of the three types of features in the following.
\begin{itemize}
    \item \textit{Categorical}: For a categorical feature field $i$, given a one-hot feature vector $x_i$, we have its embedding as $e_i=V_ix_i$, where the embedding matrix $V_i \in \mathbb{R}^{dxn}$ has vocabulary size $n$ and embedding dimension $d$.
    \item \textit{Numeric}: For a numeric feature field $j$, there are multiple choices for feature embedding: \texttt{1)} one can bucketize numeric values  into discrete features, either by designing manually (e.g., grouping age 13$\sim$19 as teenager) or through training decision trees over numeric features (e.g., GBDT~\cite{GBDT_LR}), and then embed them as categorical features; \texttt{2)} Given a \textit{normalized} scalar value $x_j$, we set its embedding as $e_j=v_jx_j$, where $v_j\in\mathbb{R}^{d}$ is the shared embedding vector of all features in field $j$; \texttt{3)} Instead of bucketizing each value into one bucket or assigning one single vector to each numeric field, one can apply AutoDis~\cite{AutoDis}, a numeric feature embedding method, to dynamically bucketizing the numeric feature and computing the embedding from a meta embedding matrix.
    \item \textit{Multi-valued}: For a multi-valued field $h$, each feature can be represented as a sequence. We obtain its embedding as $e_h=V_h[x_{h1},x_{h2},\hdots,x_{hk}]\in\mathbb{R}^{dxk}$, given $x_{hk}$ as a one-hot encoded vector of the sequence element and $k$ denoting the maximal length of the sequence. Then the embedding $e_h$ can be further aggregated to a $d$-dimensional vector, e.g., through mean/sum pooling. A further potential improvement is to apply sequential models, such as target attention in DIN~\cite{DIN} and GRU in DIEN~\cite{DIEN}, to aggregate the multi-valued behaviour sequence features.
\end{itemize}

\subsubsection{Feature Interaction}
It is straightforward to apply any classification model for CTR prediction after feature embedding. Nevertheless, for CTR prediction tasks, interactions between features (a.k.a., feature conjunctions) are central to boost classification performance. In factorization machines (FM)~\cite{FM}, inner products are shown as a simple yet effective way to capture pairwise feature interactions. Since the success of FM, a large body of research has been devoted to capturing interactions among features in different manners. Typical examples include inner product and outer product layers in PNN~\cite{PNN}, Bi-interaction in NFM~\cite{NFM}, cross network in DCN~\cite{DCN}, compressed interaction in xDeepFM~\cite{xDeepFM}, convolution in FGCNN~\cite{CNN-FeatureGen}, circular convolution in HFM~\cite{HFM}, bilinear interaction in FiBiNET~\cite{FiBiNET}, self-attention in AutoInt~\cite{autoint}, graph neural network in FiGNN~\cite{FiGNN}, hierarchical attention in InterHAt~\cite{InterHAt}, just to name a few. Furthermore, most current work investigates the way to combine both explicit and implicit feature interactions with vanilla fully-connected networks (i.e., MLPs).

% \subsubsection{Aggregation Network}
% Wide\&Deep~\cite{WideDeep} is a general learning framework that combine a wide network and a deep network for joint training. The wide network captures low-order feature interactions and is able to memorize the frequently co-occurring features. The deep network models complex high-order feature interactions through low-dimensional dense embeddings, which generalize better to unseen feature combinations. Through joint training of such two-branch networks, it achieves both memorization and generalization in one
% model~\cite{WideDeep}. Due to its success, many following models take this framework to combine two different networks (e.g., shallow and deep networks~\cite{DeepFM}, explicit and implicit feature interactions~\cite{DCN,xDeepFM,autoint}) to bring additional improvement over a single branch network. Notably, whereas the Wide\&Deep framework mostly takes the simple summation of two networks as the final output, the recent work~\cite{NON} proposes a non-linear DNN to fuse different branch networks to enhance the interactions among them. 

\subsubsection{Loss Function}
The binary cross-entropy loss is widely used in CTR prediction tasks, which is defined as follows:
\begin{equation}
    \mathcal{L}= - \frac{1}{N}\sum_{\mathbb{D}} \Big(y log{\hat{y}} + (1-y)log(1-\hat{y})\Big)~,
    \label{equ:bce}
\end{equation}
where $\mathbb{D}$ is the training set with $N$ samples. $y$ and $\hat{y}$ denote the ground truth and the estimated click probability, respectively. We define $\hat{y}=\sigma(\phi(x))$, where $\phi(x)$ represents the model function given input features $x$ and $\sigma(\cdot)$ is the sigmoid function to map $\hat y$ to $[0,1]$. The core of CTR prediction modeling lies in how to construct the model $\phi(x)$ and learn its parameters from training data.

\subsection{Representative Models}\label{sec:ctr}
In this section, we summarize the representative models that we have evaluated and benchmarked in this work. Note that although we enumerate only a part of the existing models above, they have covered a wide spectrum of studies on CTR prediction.

\subsubsection{Shallow Models}
It is common that industrial CTR prediction tasks have large-scale data. Therefore, shallow models have been in widespread use due to their simplicity and efficiency. Even today, LR~\cite{LR} and FM~\cite{FM} are still two strong baseline models deployed in industry. We describe the shallow models as follows:

\begin{itemize}[leftmargin=*]
\item\textbf{LR}. Logistic regression (LR) is a simple baseline model for CTR prediction~\cite{LR}. With the online learning algorithm, FTRL~\cite{FTRL}, proposed by Google, LR has been widely adopted in industry.

\item\textbf{FM}. While LR fails to capture non-linear feature interactions, Rendle et al. propose factorization machine (FM)~\cite{FM} that embeds features into dense vectors and models pairwise feature interactions as inner products of the corresponding embedding vectors. Notably, FM also has a linear time complexity in terms of the number of features.

\item\textbf{FFM}. Field-aware factorization machine (FFM)~\cite{FFM} is an extension of FM that considers field information for feature interactions. It was a winner model in several Kaggle contests on CTR prediction. 

\item\textbf{HOFM}. Since FM only captures second-order feature interactions, HOFM~\cite{HighFM} aims to extend FM to higher-order factorization machines. However, it results in exponential feature combinations that consume huge memory and take a long running time.

\item\textbf{FwFM}. Recently, Pan et al.~\cite{FwFM} extends FM by considering the field-wise weights of features interactions. Compared with FFM, it reports comparable performance but uses much fewer model parameters. 

\item\textbf{LorentzFM}. LorentzFM~\cite{LorentzFM} has recently been proposed to embed features into a hyperbolic space and model feature interactions via triangle inequality of Lorentz distance. 

%\item\textbf{FmFM}. As an improvement of FwFM~\cite{FwFM}, FmFM~\cite{FmFM} employs a field-wise weighting matrix to model the field information of features interactions. In spirit, it equals to the bilinear interactions proposed in FiBiNET~\cite{FiBiNET}.
\end{itemize}

%%%%%%%%%%%%%%%%%%%%%%%%%%%%%%%%%%%%%%%%%%%
\begin{table*}[!t]
\renewcommand\arraystretch{1.0}
% \setlength{\belowcaptionskip}{10pt}
% \setlength{\tabcolsep}{3.7pt}
%\small
\centering
\caption{Reproducibility requirements met by existing studies (\cmark ~| \textminus ~| \xmark ~means totally | partially | not met, respectively)}\label{tab:reproducibility}
\vspace{-1ex}
\begin{tabular}{l|c|c|c|c|c|c|c|c|c|c}
\hline
Reproducibility requirements & xDeepFM & FGCNN & AutoInt+ & FiGNN & ONN & FiBiNET & LorentzFM & AFN+ & InterHAt  & Ours \\ \hline
Data preprocessing & \xmark & \xmark  & \cmark & \xmark  & \xmark  & \xmark  & \xmark  & \cmark  & \cmark &  \cmark\\
Model source code & -- & -- & \cmark & \cmark & -- & -- & \xmark & \cmark & \cmark & \cmark \\
Model hyper-parameters & \cmark & \cmark & \cmark & \cmark & \cmark & \cmark & \cmark & \cmark & \cmark & \cmark \\
Baseline source code & \xmark & \xmark & \xmark & \xmark & \xmark & \xmark & \xmark & \xmark & \xmark & \cmark \\ 
Baseline hyper-parameters & \cmark & \cmark & \textminus  & \textminus  & \xmark & \textminus  & \textminus  & \textminus  & \xmark & \cmark \\
%Running logs & \xmark & \xmark & --  & --  & \xmark & \xmark  & \xmark  & \xmark & \xmark & \xmark & \cmark \\
\hline
\end{tabular}
\vspace{1ex}
\end{table*}
%%%%%%%%%%%%%%%%%%%%%%%%%%%%%%%%%%%%%%%%%%%

\subsubsection{Deep Models}
Nowadays, deep neural networks have been widely studied and applied for CTR prediction. Compared with shallow models, deep models are more powerful in capturing sophisticated high-order feature interactions with non-linear activation functions, which usually yield better performance. Yet, efficiency has become the major bottleneck to scale deep models in practice3.

\begin{itemize}[leftmargin=*]
\item\textbf{DNN}. DNN is a straightforward deep model reported in \cite{DNN}, which applies a fully-connected network (termed DNN) after the concatenation of feature embeddings for CTR prediction. 
\item\textbf{CCPM}. CCPM~\cite{CCPM} reports the first attempt to use convolution for CTR prediction, where feature embeddings are aggregated hierarchically through convolution networks.
\item\textbf{Wide\&Deep}. Wide\&Deep~\cite{WideDeep} is a general learning framework proposed by Google that combines a wide (or shallow) network and deep network to achieve the advantages of both.
\item\textbf{IPNN}. PNN~\cite{PNN} is a product-based network that feeds the inner (or outer) products of features embeddings as the input of DNN. Due to the huge memory requirement of pairwise outer products, we choose the inner product version, IPNN.
\item\textbf{DeepCross}. Inspired by residual networks, ~\cite{DeepCross} propose deep crossing to add residual connections between layers of DNNs.
\item\textbf{NFM}. Similar to PNN, NFM~\cite{NFM} proposes a Bi-interaction layer that pools the pairwise feature interactions to a vector and then feed it to a DNN for CTR prediction.
\item\textbf{AFM}. Instead of treating all feature interactions equally as in FM, AFM~\cite{AFM} learns the weights of feature interactions via attentional networks. Different from FwFM, AFM adjusts the weights dynamically according to the input data sample. 
\item\textbf{DeepFM}. DeepFM~\cite{DeepFM} is an extension of Wide\&Deep that substitutes LR with FM to explicitly model second-order feature interactions. 
\item\textbf{DCN}. In DCN~\cite{DCN}, a cross network is proposed to perform high-order feature interactions in an explicit way. In addition, it also integrates a DNN network following the Wide\&Deep framework. 
\item\textbf{xDeepFM}. While high-order feature interactions modeled by DCN are bit-wise,  xDeepFM~\cite{xDeepFM} proposes to capture high-order feature interactions in a vector-wise way via a compressed interaction network (CIN).
\item\textbf{HFM+}. HFM~\cite{HFM} proposes holographic representation and computes compressed outer products via circular convolution to model pairwise feature interactions. HFM+ further integrates a DNN network with HFM.
\item\textbf{FGCNN}. FGCNN~\cite{CNN-FeatureGen} applies convolution networks and recombination layers to generate additional combinatorial features to enrich existing feature representations.
\item\textbf{AutoInt+}. AutoInt~\cite{autoint} leverages self-attention networks to learn high-order features interactions. AutoInt+ integrates AutoInt with a DNN network.
\item\textbf{FiGNN}. FiGNN~\cite{FiGNN} leverages the message passing mechanism of graph neural networks to learn high-order features interactions. 
\item\textbf{ONN}. ONN (a.k.a., NFFM)~\cite{NFFM} is a model built on FFM. It feeds the interaction outputs from FFM to a DNN network for CTR prediction. 
\item\textbf{FiBiNET}. FiBiNET~\cite{FiBiNET} leverages squeeze-excitation network to capture important features, and proposes bilinear interactions to enhance feature interactions.
\item\textbf{AFN+}. AFN~\cite{AFN} applies logarithmic transformation layers to learn adaptive-order feature interactions. AFN+ further integrates AFN with a DNN network.
\item\textbf{InterHAt}. InterHAt~\cite{InterHAt} employs hierarchical attention networks to model high-order feature interactions in an efficient manner. 
\end{itemize}

%% file: sections/4_experiment.tex
\section{CTR Benchmarking}\label{sec:exp}
In this section, we first summarize the key reproducibility requirements, and then present our evaluation protocols and open-source toolkit for benchmarking CTR prediction models. Finally, we report and analyze our benchmarking results. 

\subsection{Reproducibility Requirements}
CTR prediction has been widely studied in recent years. Unfortunately, there is a lack of rigour on reproducibility in many studies. This may lead to non-reproducible and inconsistent results reported in existing studies, thereby hindering the development and evolution of new techniques in this field. In this work, we highlight the following five key requirements to ensure reproducible research. Yet, many current studies fail to meet all of these reproduciblility requirements, as shown in Table~\ref{tab:reproducibility}. Note that we use \cmark ~| \textminus ~| \xmark ~to denote whether each requirement is totally, partially, or not met, respectively. It also indicates whether the artifacts are totally, partially, or not available for reproducing each step. For example, "--" indicates that FiBiNET has only unofficial model source code.  

\begin{itemize}[leftmargin=*]
\item\textbf{Data preprocessing}: Most work splits the training, validation, and testing data randomly, but others often cannot repeat the same data splitting due to the lack of scripts or random seeds used. Even more, some preprocessing details (e.g., how to handle numeric features and what threshold is used to filter rare categorical features) may be missing or incomplete. In such cases, researchers have to perform their own data splitting and preprocessing, leading to uncomparable results. It is worth noting that the authors of AutoInt~\cite{autoint}, AFN~\cite{AFN}, and InterHAt~\cite{InterHAt} have made a good starting point to share the data processing code or the preprocessed data.
\item\textbf{Model source code}: In the spirit of open source, some studies have released the source code of their models on GitHub. Some popular models also have unofficial implementations available online from third-party libraries (e.g., DeepCTR~\cite{DeepCTR}). But in many cases, we found that the source code is not ready for reproducibility studies, because it may either lack the training code (e.g., loading data, early stopping, etc.) or miss some key hyper-parameters on the given dataset.
\item\textbf{Model hyper-parameters}: Most studies specify the detailed hyper-parameters of their own models in the papers. But without access to the original data preprocessed by the authors, it is inappropriate for others to use the same hyper-parameters on new data splits. It may only attain sub-optimal performance and needs to be retuned. Such practices lead to inconsistent results among existing papers.
\item\textbf{Baseline source code}: Many studies report the details of their own models, but fail to clarify how they apply the baseline models. We noticed that existing studies seldom open source the baseline models, or tell which implementation is used for comparison. Model performance depends heavily on the quality of their code implementations. Bad implementations may introduce biases and make unfair model comparisons. However, this aspect is often overlooked by existing studies, making their performance improvements difficult to reproduce. % FmFM~\cite{FmFM} has done a good job in this aspect.  
\item\textbf{Baseline hyper-parameters}: It is desirable to exhaustively tune the hyper-parameters of baseline models to fairly compare model performance. Yet, this has not been guaranteed due to the lack of open benchmarking. Most existing studies usually report inconsistent results of the same baseline models due to their unknown data preprocessing and baseline implementations. 
\end{itemize}

To enable reproducible research and fairness of comparison, in this work, we aim to set up a standardized benchmarking pipeline as well as provide the most comprehensive open benchmarking results for CTR prediction.

% Thus a standardized benchmark and common evaluation methodology for current models are highly demanded by both academic research and industrial applications. 

\subsection{Evaluation Protocol}\label{sec:protocol}

\subsubsection{Datasets} 
In this work, we mainly use two real-world datasets for our evaluation: Criteo~\cite{Criteo} and Avazu~\cite{Avazu}. Both of them are open datasets released by two leading ad companies, and have been widely used in the previous work (e.g.,~\cite{DeepFM, DCN, xDeepFM, autoint}). We choose them because they are collected or sampled from real click logs in production, and both have tens of millions of samples, making the benchmarking results meaningful to industrial practitioners. Table \ref{tab:Data Statis} summarizes the data statistics information. We further supplement benchmarking results on more datasets on the FuxiCTR leaderboard.

%%%%%%%%%%%%%%%%%%%%%%%%%%%%%%%%%%%%%%%%%%%%%%%%%%%%%%%%%%%%%%%%%%%%
\begin{table}[!t]
%\small
\centering
\caption{Dataset statistics}\label{tab:Data Statis} 
\begin{tabular}{c||c|c|c|c}
\hline
{Dataset}& \#Instances & \#Fields & \#Features & \%Positives \\
\hline 
 Criteo  & 46M & 39 & 5.55M & 26\% \\
\hline
  Avazu & 40M & 24 & 8.37M & 17\% \\
\hline
\end{tabular}
% \vspace{-2ex}
\end{table}
%%%%%%%%%%%%%%%%%%%%%%%%%%%%%%%%%%%%%%%%%%%%%%%%%%%%%%%%%%%%%%%%%%%%

\subsubsection{Data splitting} As with most of the existing studies, we follow them to randomly split both Criteo and Avazu into 8:1:1 as the training set, validation set, and test set, respectively. To make it exactly reproducible and easy to compare with existing work, we reuse the code provided by AutoInt~\cite{autoint} and control the random seed (i.e., seed=$2018$) for splitting. We mark these two data splits as \textit{Criteo\_x4} and \textit{Avazu\_x4}, respectively.

\subsubsection{Data preprocessing} We mostly follow the same steps as in~\cite{autoint} to preprocess the features. Moreover, we make some modifications and fix a defect in~\cite{autoint} to improve the benchmark results.

\begin{itemize}[leftmargin=*]
\item\textbf{Criteo}. The Criteo dataset consists of ad click data over a week. It comprises 26 categorical feature fields and 13 numerical feature fields. We create two different evaluation settings, denoted as \textit{Criteo\_x4\_001} and \textit{Criteo\_x4\_002}. Concretely, in {Criteo\_x4\_001}, instead of normalizing numeric values as in~\cite{autoint}, we follow the winner solution of the Criteo contest to discretize each numeric value $x$ to $\lfloor\log^2(x)\rfloor$, if $x > 2$; $x=1$ otherwise, which yields much better performance. For categorical features, we replace infrequent features (min\_count=10) with a default "OOV" token. We also fix the feature embedding dimension to 16 in this setting. {Criteo\_x4\_002} differs in that we set min\_count=2 for categorical features, and the embedding dimension equals to 40 after tuning. 

%% Introduce Avazu and its preparation
\item\textbf{Avazu}. Avazu contains 10 days of click logs. It has a total of 23 fields with categorical features including app id, app category, device id, etc. We also create two evaluation settings, namely \textit{Avazu\_x4\_001} and \textit{Avazu\_x4\_002}. In Avazu\_x4\_001, we remove the \texttt{id} field that has a unique value in each data sample, which should be useless for CTR prediction. But it is retained in~\cite{autoint}, resulting in a defect. In addition, we transform the timestamp field into three new fields: hour, weekday, and is\_weekend. For all categorical features, we replace infrequent features (min\_count=2) with a default "OOV" token. We further fix the feature embedding dimension to 16 as in~\cite{autoint}. Avazu\_x4\_002 is different in that we set min\_count=1 and fix the embedding dimension to 40. We emphasize that for Avazu\_x4, we set a small threshold because it indeed results in much better performance than the original setting (min\_count=10) in~\cite{autoint}. Yet, we note that the relative comparisons among different models in our benchmark are fair, since all the models are in the same embedding size. Our benchmark strives to raise the bar of baselines when make model comparisons.
\end{itemize}

\subsubsection{Evaluation metrics} 
We employ two most commonly-used metrics, AUC and logloss, for benchmarking.
\begin{itemize}[leftmargin=*]
\item\textbf{AUC} (i.e., Area Under the ROC Curve) is a common metric to measure the probability that a randomly chosen positive sample is ranked higher than a randomly chosen negative sample. Higher AUC indicates better CTR prediction performance.

\item\textbf{Logloss}, also known as the logistic loss or binary cross-entropy loss, is defined as Equation~\ref{equ:bce}. Lower logloss indicates better CTR prediction performance.
\end{itemize}

It is worth noting that, when considering a large user base, an improvement of AUC at 0.001 level is generally considered as practically significant for an industrial CTR prediction task, as indicated by several existing studies from Google~\cite{WideDeep,DCN}, Microsoft~\cite{EnsembleBing}, and Huawei~\cite{DeepFM}. Note that although the metric gAUC~\cite{DIN} is also widely employed in practice, we cannot report the results because user\_id is missing in both Criteo and Avazu datasets. 

\subsubsection{Benchmarking toolkit} 
While many open-source projects exist for CTR prediction, they mostly implement a few models in an ad-hoc manner and lack a complete workflow for benchmarking. In particular, DeepCTR~\cite{DeepCTR} provides a good package with uniform implementations of many CTR prediction models. Nevertheless, our benchmarking needs a complete workflow, including data preprocessing, batch loading, model training, early stopping,  learning rate decay, hyper-parameter search, and most importantly, seeding and logging for reproducibility. 

To this end, we develop the open-source FuxiCTR library\footnote{FuxiCTR Library: \textcolor{blue!70!black}{\url{https://github.com/reczoo/FuxiCTR}}} for CTR prediction, providing stunning features about configurability, tunability, and reproducibility. Since its inception in 2018, the project has evolved for over three years. The framework of FuxiCTR consists of the following parts: 1) The data preprocessing part reads the raw data from CSV files, transforms all numeric, categorical, and sequential features, and outputs the transformed data into HDF5 files. 2) Tens of models have already been implemented in a uniform way in Pytorch. 3) The training part is implemented to read batches of data, compute forward and backward passes, and perform learning rate decay and early stopping if necessary. 4) The seeding and logging utilities are also specially designed for reproducibility, recording detailed running logs (including the used hyper-parameters) for each benchmarking experiment. 4) The hyper-parameter tuning part provides a configurable interface to allow grid search of hyper-parameters specified by users. We integrate all these parts as a complete benchmarking framework, allowing researchers to easily reuse our code, build new models, or adding new datasets. The ultimate goal is to provide an easy-to-use software package for reproducible research on CTR prediction.

\subsubsection{Training details and hyper-parameters tuning} 
During training, we apply the Reduce-LR-on-Plateau scheduler by default to reduce learning rate by a factor of $10$ when the given metric stops improving. To avoid overfitting, early stopping is employed when the metric on the validation set stops improving in 2 or 3 consecutive epochs. The default learning rate is $1.e-3$. The batch size is initially set to 10000 and then decreases gradually using [5000, 2000, 1000] if an OOM error occurs in GPUs. We found that for CTR prediction models trained on millions of samples, using a large batch size usually makes the model run much faster and attains better results. Given the large number of features, feature embeddings usually take up most of the model parameters. For fairness of comparison, we fix the embedding size to $16$ or $40$ in two seperate settings. We also found that regularization weight makes a large effect on the model performance. Thus, we carefully tuned it in the range of 0$\sim$1, usually at a multiplicative ratio of 10x. The model size (e.g., number of layers and units) is highly data-dependent, so we exhaustively tune these hyper-parameters. We also carefully tune some other hyper-parameters (e.g., whether use bath normalization), if any, to attain the best result for each model. To avoid exponential combination space, we usually tune important hyper-parameters first and then the other ones group by group. On average, we run 73 experiments for each model to obtain the best result. All the experiments were run on a shared GPU cluster with P100 GPUs, each with 16GB GPU memory.

\subsubsection{Reproducibility} 
For reproduciblility, we keep the md5sum values of each data split. We explicitly set the random seed for each experiment and record the data settings and model hyper-parameters into configure files. In particular, we choose to implement the models in Pytorch, since it offers better ability than Tensorflow to avoid non-determinism when running models on a GPU device. We have open source the benchmarking code together with all the evaluation settings and results to the community to foster more reproducible research in the future.

%%%%%%%%%%%%%%%%%%%%%%%%%%%%%%%%%%%%%%%%%%%%%%%%%%%%%%%%%%
%https://www.tablesgenerator.com/latex_tables
\begin{table*}[]
\centering
\renewcommand\arraystretch{1.0}
\setlength{\tabcolsep}{3.2pt}
% \small
% \vspace{-1ex}
\caption{Benchmarking results on Criteo and Avazu. We highlight the
top-5 results in each dataset setting (Logloss $\times10^{-2}$).}\label{tab:benchmarking}
% \vspace{-3ex}
%(Bold numbers indicate the best results in each column; the underscored numbers show the results that have not been successfully reproduced)

\begin{tabular}{ccccllccllccl}
\hline
\multicolumn{13}{c}{\cellcolor[HTML]{fff8f8}{\textbf{Criteo}}} \\ \hline
\multicolumn{1}{c|}{\multirow{2}{*}{Year}} & \multicolumn{1}{c|}{\multirow{2}{*}{Model}} & \multicolumn{2}{c|}{Best Reported} & \multicolumn{4}{c|}{Criteo\_x4 (dim=16)}                   & \multicolumn{5}{c}{Criteo\_x4 (dim=40)}            \\ \cline{3-13} 
\multicolumn{1}{c|}{}                      & \multicolumn{1}{c|}{}                       & Logloss & \multicolumn{1}{c|}{AUC(\%)} & Logloss & \multicolumn{1}{l|}{ AUC(\%)} & \multicolumn{1}{c|}{\#Params} & \multicolumn{1}{c|}{\#Runs} & Logloss & \multicolumn{1}{l|}{ AUC(\%)} & \multicolumn{1}{c|}{\#Params} & \multicolumn{1}{c|}{\#Runs} & Time$\times$Epochs \\ \hline
\multicolumn{1}{c|}{2007} & \multicolumn{1}{c|}{LR} & {44.74} & \multicolumn{1}{c|}{78.58} & 45.68 & \multicolumn{1}{l|}{ 79.34} & \multicolumn{1}{c|}{0.9M} & \multicolumn{1}{c|}{6} & \multicolumn{1}{l}{ 45.66} & \multicolumn{1}{l|}{ 79.36} & \multicolumn{1}{c|}{5.5M} & \multicolumn{1}{c|}{12} & 7m $\times$ 12 \\
\multicolumn{1}{c|}{2010} & \multicolumn{1}{c|}{FM} & 44.64 & \multicolumn{1}{c|}{79.33} &44.31 & \multicolumn{1}{l|}{ 80.86} & \multicolumn{1}{c|}{15.5M} & \multicolumn{1}{c|}{9} & \multicolumn{1}{l}{ 44.45} & \multicolumn{1}{l|}{ 80.78} & \multicolumn{1}{c|}{227.5M} & \multicolumn{1}{c|}{20} & 18m $\times$ 5 \\
\multicolumn{1}{c|}{2015} & \multicolumn{1}{c|}{CCPM} & — & \multicolumn{1}{c|}{—} & 44.15 & \multicolumn{1}{l|}{ 81.04} & \multicolumn{1}{c|}{1.7M} & \multicolumn{1}{c|}{9} & \multicolumn{1}{l}{ 44.40} & \multicolumn{1}{l|}{ 80.77} & \multicolumn{1}{c|}{222.0M} & \multicolumn{1}{c|}{24} & 2h33m $\times$ 1 \\
\multicolumn{1}{c|}{2016} & \multicolumn{1}{c|}{HOFM} & 45.08 & \multicolumn{1}{c|}{80.05} & 44.11 & \multicolumn{1}{l|}{ 81.07} & \multicolumn{1}{c|}{30.1M} & \multicolumn{1}{c|}{8} & \multicolumn{1}{l}{ 44.04} & \multicolumn{1}{l|}{ 81.15} & \multicolumn{1}{c|}{255.3M} & \multicolumn{1}{c|}{17} & 1h42m $\times$ 26 \\
\multicolumn{1}{c|}{2016} & \multicolumn{1}{c|}{FFM} & 45.25 & \multicolumn{1}{c|}{80.01} & 44.07 & \multicolumn{1}{l|}{ 81.13} & \multicolumn{1}{c|}{139.3M} & \multicolumn{1}{c|}{18} & \multicolumn{1}{l}{ 44.09} & \multicolumn{1}{l|}{ 81.11} & \multicolumn{1}{c|}{638.2M} & \multicolumn{1}{c|}{20} & 3h59m $\times$ 5 \\
\multicolumn{1}{c|}{2016} & \multicolumn{1}{c|}{DNN} & 44.91 & \multicolumn{1}{c|}{79.93} & 43.80 & \multicolumn{1}{l|}{ 81.40} & \multicolumn{1}{c|}{19.2M} & \multicolumn{1}{c|}{51} & \multicolumn{1}{l}{ 44.07} & \multicolumn{1}{l|}{ 81.12} & \multicolumn{1}{c|}{226.5M} & \multicolumn{1}{c|}{64} & 13m $\times$ 2 \\
\multicolumn{1}{c|}{2016} & \multicolumn{1}{c|}{WideDeep} & 44.53 & \multicolumn{1}{c|}{80.62} & \textbf{43.77(5)} & \multicolumn{1}{c|}{\textbf{81.42(5)}} &  \multicolumn{1}{c|}{20.1M} & \multicolumn{1}{c|}{30} & \multicolumn{1}{l}{ 43.89} & \multicolumn{1}{l|}{ 81.29} & \multicolumn{1}{c|}{231.1M} & \multicolumn{1}{c|}{80} & 11m $\times$ 4  \\
\multicolumn{1}{c|}{2016} & \multicolumn{1}{c|}{IPNN} & 45.32 & \multicolumn{1}{c|}{80.38} & 43.78 & \multicolumn{1}{l|}{\textbf{81.42(5)}} & \multicolumn{1}{c|}{16.9M} & \multicolumn{1}{c|}{24} & \multicolumn{1}{l}{ 43.88} & \multicolumn{1}{l|}{ 81.32} & \multicolumn{1}{c|}{258.5M} & \multicolumn{1}{c|}{38} & 32m $\times$ 2 \\
\multicolumn{1}{c|}{2016} & \multicolumn{1}{c|}{DeepCross} & 44.25 & \multicolumn{1}{c|}{80.09} & 43.84 & \multicolumn{1}{l|}{ 81.35} & \multicolumn{1}{c|}{17.1M} & \multicolumn{1}{c|}{30} & \multicolumn{1}{c}{\textbf{43.80(5)}} & \multicolumn{1}{l|}{ 81.39} & \multicolumn{1}{c|}{284.4M} & \multicolumn{1}{c|}{138} & 26m $\times$ 3 \\
\multicolumn{1}{c|}{2017} & \multicolumn{1}{c|}{AFM} & 45.41 & \multicolumn{1}{c|}{79.65} & 44.55 & \multicolumn{1}{l|}{ 80.60} & \multicolumn{1}{c|}{15.5M} & \multicolumn{1}{c|}{15} & \multicolumn{1}{l}{ 44.43} & \multicolumn{1}{l|}{ 80.73} & \multicolumn{1}{c|}{227.5M} & \multicolumn{1}{c|}{10} & 31m $\times$ 22 \\
\multicolumn{1}{c|}{2017} & \multicolumn{1}{c|}{NFM} & 45.37 & \multicolumn{1}{c|}{79.68} & 44.24 & \multicolumn{1}{l|}{ 80.93} & \multicolumn{1}{c|}{17.5M} & \multicolumn{1}{c|}{24} & \multicolumn{1}{l}{ 44.43} & \multicolumn{1}{l|}{ 80.72} & \multicolumn{1}{c|}{229.6M} & \multicolumn{1}{c|}{64} & 18m $\times$ 2 \\
\multicolumn{1}{c|}{2017} & \multicolumn{1}{c|}{DeepFM} & 44.45 & \multicolumn{1}{c|}{80.85} & \textbf{43.76(2)} & \multicolumn{1}{c|}{\textbf{81.43(3)}} & \multicolumn{1}{c|}{20.1M} & \multicolumn{1}{c|}{30} & \multicolumn{1}{c}{\textbf{43.78(2)}} & \multicolumn{1}{c|}{\textbf{81.41(3)}} & \multicolumn{1}{c|}{229.1M} & \multicolumn{1}{c|}{128} & 18m $\times$ 9 \\
\multicolumn{1}{c|}{2017} & \multicolumn{1}{c|}{DCN} & 44.19 & \multicolumn{1}{c|}{80.67} & \textbf{43.76(2)} & \multicolumn{1}{c|}{\textbf{81.44(2)}} & \multicolumn{1}{c|}{19.2M} & \multicolumn{1}{c|}{81} & \multicolumn{1}{c}{\textbf{43.78(2)}} & \multicolumn{1}{c|}{\textbf{81.41(3)}} & \multicolumn{1}{c|}{245.1M} & \multicolumn{1}{c|}{544} & 14m $\times$ 9 \\
\multicolumn{1}{c|}{2018} & \multicolumn{1}{c|}{FwFM} & — & \multicolumn{1}{c|}{—} & 44.08 & \multicolumn{1}{l|}{ 81.12} & \multicolumn{1}{c|}{15.5M} & \multicolumn{1}{c|}{18} & \multicolumn{1}{l}{ 44.19} & \multicolumn{1}{l|}{ 80.98} & \multicolumn{1}{c|}{222.0M} & \multicolumn{1}{c|}{14} & 19m $\times$ 1 \\
\multicolumn{1}{c|}{2018} & \multicolumn{1}{c|}{xDeepFM} & 44.18 & \multicolumn{1}{c|}{80.91} & \textbf{43.76(2)} & \multicolumn{1}{c|}{\textbf{81.43(3)}} &  \multicolumn{1}{c|}{20.5M} & \multicolumn{1}{c|}{106} & \multicolumn{1}{c}{\textbf{43.75(1)}} & \multicolumn{1}{c|}{\textbf{81.44(1)}} & \multicolumn{1}{c|}{232.3M} & \multicolumn{1}{c|}{156} & 1h9m $\times$ 8 \\
\multicolumn{1}{c|}{2019} & \multicolumn{1}{c|}{HFM+} & — & \multicolumn{1}{c|}{—} & 43.92 & \multicolumn{1}{l|}{ 81.27} & \multicolumn{1}{c|}{29.3M} & \multicolumn{1}{c|}{34} & \multicolumn{1}{l}{ 43.91} & \multicolumn{1}{l|}{ 81.27} & \multicolumn{1}{c|}{260.2M} & \multicolumn{1}{c|}{74} & 1h13m $\times$ 2 \\
\multicolumn{1}{c|}{2019} & \multicolumn{1}{c|}{FGCNN} & — & \multicolumn{1}{c|}{—} & 	43.98 & \multicolumn{1}{l|}{ 81.21} & \multicolumn{1}{c|}{61.1M} & \multicolumn{1}{c|}{56} & \multicolumn{1}{l}{  43.81} & \multicolumn{1}{c|}{\textbf{81.42(2)}} & \multicolumn{1}{c|}{317.4M} & \multicolumn{1}{c|}{87} & 3h9m $\times$ 6 \\
\multicolumn{1}{c|}{2019} & \multicolumn{1}{c|}{AutoInt+} & 44.34 & \multicolumn{1}{c|}{80.83} & 43.90 & \multicolumn{1}{l|}{ 81.32} & \multicolumn{1}{c|}{20.2M} & \multicolumn{1}{c|}{75} & \multicolumn{1}{l}{ 43.85} & \multicolumn{1}{l|}{ 81.34} & \multicolumn{1}{c|}{285.4M} & \multicolumn{1}{c|}{120} & 16m $\times$ 1 \\
\multicolumn{1}{c|}{2019} & \multicolumn{1}{c|}{FiGNN} & 44.53 & \multicolumn{1}{c|}{80.62} & 43.83 & \multicolumn{1}{l|}{ 81.38} & \multicolumn{1}{c|}{14.7M} & \multicolumn{1}{c|}{45} & \multicolumn{1}{c}{\textbf{43.79(4)}} & \multicolumn{1}{c|}{\textbf{81.41(3)}} &\multicolumn{1}{c|}{222.7M} & \multicolumn{1}{c|}{40} & 1h40m $\times$ 14 \\
\multicolumn{1}{c|}{2019} & \multicolumn{1}{c|}{ONN} & {{43.58}} & \multicolumn{1}{c|}{{81.23}} & \textbf{43.72(1)}& \multicolumn{1}{c|}{\textbf{81.48(1)}} & \multicolumn{1}{c|}{287.2M} & \multicolumn{1}{c|}{30} & \multicolumn{1}{l}{ 43.81} & \multicolumn{1}{l|}{ \textbf{81.41(3)}} &\multicolumn{1}{c|}{436.7M} & \multicolumn{1}{c|}{108} & 2h13m $\times$ 8 \\
\multicolumn{1}{c|}{2019} & \multicolumn{1}{c|}{FiBiNET} & 44.23 & \multicolumn{1}{c|}{81.03} & 43.87 & \multicolumn{1}{l|}{ 81.31} &  \multicolumn{1}{c|}{71.1M} & \multicolumn{1}{c|}{96} & \multicolumn{1}{l}{ 43.86} & \multicolumn{1}{l|}{ 81.34} & \multicolumn{1}{c|}{482.5M} & \multicolumn{1}{c|}{78} & 2h21m $\times$ 2 \\
\multicolumn{1}{c|}{2020} & \multicolumn{1}{c|}{LorentzFM} & — & \multicolumn{1}{c|}{—} & 44.34 & \multicolumn{1}{l|}{ 80.83} & \multicolumn{1}{c|}{14.6M} & \multicolumn{1}{c|}{14} & \multicolumn{1}{l}{ 44.13} & \multicolumn{1}{l|}{ 81.05} & \multicolumn{1}{c|}{222.0M} & \multicolumn{1}{c|}{8} & 2h12m $\times$ 20 \\
\multicolumn{1}{c|}{2020} & \multicolumn{1}{c|}{AFN+} & 44.51 & \multicolumn{1}{c|}{80.74} & 43.84 & \multicolumn{1}{l|}{ 81.38} &  \multicolumn{1}{c|}{56.0M} & \multicolumn{1}{c|}{79} & \multicolumn{1}{l}{ 43.87} & \multicolumn{1}{l|}{ 81.34} &\multicolumn{1}{c|}{238.1M} & \multicolumn{1}{c|}{114} & 38m $\times$ 6 \\
\multicolumn{1}{l|}{2020} & \multicolumn{1}{l|}{InterHAt} & 45.77 & \multicolumn{1}{c|}{78.45} & \multicolumn{1}{l}{ 44.14} & \multicolumn{1}{l|}{ 81.04} & \multicolumn{1}{c|}{15.6M} & \multicolumn{1}{c|}{106} & \multicolumn{1}{l}{ 44.01} & \multicolumn{1}{l|}{ 81.17} & \multicolumn{1}{c|}{222.1M} & \multicolumn{1}{c|}{90} & 18m $\times$ 20 \\ 
\hline
 &  &  &  &  &  & & & &  &  &  &  \\ \hline
 \multicolumn{13}{c}{\cellcolor[HTML]{fff8f8}{\textbf{Avazu}}} \\ \hline
\multicolumn{1}{c|}{\multirow{2}{*}{Year}} & \multicolumn{1}{c|}{\multirow{2}{*}{Model}} & \multicolumn{2}{c|}{Best Reported} & \multicolumn{4}{c|}{Avazu\_x4 (dim=16)}                   & \multicolumn{5}{c}{Avazu\_x4 (dim=40)}            \\ \cline{3-13} 
\multicolumn{1}{c|}{}                      & \multicolumn{1}{c|}{}                       & Logloss & \multicolumn{1}{c|}{AUC(\%)} & Logloss & \multicolumn{1}{l|}{ AUC(\%)} & \multicolumn{1}{c|}{\#Params} & \multicolumn{1}{c|}{\#Runs} & Logloss & \multicolumn{1}{l|}{ AUC(\%)} & \multicolumn{1}{c|}{\#Params} & \multicolumn{1}{c|}{\#Runs} & Time$\times$Epochs \\ \hline
\multicolumn{1}{c|}{2007} & \multicolumn{1}{c|}{LR} & 38.68 & \multicolumn{1}{c|}{76.76} & 38.15 & \multicolumn{1}{l|}{ 77.75} & \multicolumn{1}{c|}{3.8M} & \multicolumn{1}{c|}{10} & \multicolumn{1}{l}{ 37.99} & \multicolumn{1}{l|}{ 78.04} & \multicolumn{1}{c|}{8.4M} & \multicolumn{1}{c|}{90} & 5m $\times$ 26 \\
\multicolumn{1}{c|}{2010} & \multicolumn{1}{c|}{FM} & 37.40 & \multicolumn{1}{c|}{77.93} & 	37.54 & \multicolumn{1}{l|}{ 78.87} & \multicolumn{1}{c|}{63.8M} & \multicolumn{1}{c|}{25} & \multicolumn{1}{l}{ 37.36} & \multicolumn{1}{l|}{ 79.09} & \multicolumn{1}{c|}{343.3M} & \multicolumn{1}{c|}{187} & 18m $\times$ 1 \\
\multicolumn{1}{c|}{2015} & \multicolumn{1}{c|}{CCPM} & 38.00 & \multicolumn{1}{c|}{78.12} & 37.45 & \multicolumn{1}{l|}{ 78.92} & \multicolumn{1}{c|}{60.6M} & \multicolumn{1}{c|}{70} & \multicolumn{1}{l}{ 37.21} & \multicolumn{1}{l|}{ 79.32} & \multicolumn{1}{c|}{335.0M} & \multicolumn{1}{c|}{24} & 57m $\times$ 1 \\
\multicolumn{1}{c|}{2016} & \multicolumn{1}{c|}{HOFM} & 37.56 & \multicolumn{1}{c|}{77.01} & 37.54 & \multicolumn{1}{l|}{ 78.91} & \multicolumn{1}{c|}{123.8M} & \multicolumn{1}{c|}{10} & \multicolumn{1}{l}{ 37.33} & \multicolumn{1}{l|}{ 79.14} & \multicolumn{1}{c|}{385.2M} & \multicolumn{1}{c|}{31} & 1h47m $\times$ 1 \\
\multicolumn{1}{c|}{2016} & \multicolumn{1}{c|}{FFM} & 37.81 & \multicolumn{1}{c|}{78.31} & 37.20 & \multicolumn{1}{l|}{ 79.31} &  \multicolumn{1}{c|}{693.9M} & \multicolumn{1}{c|}{19} & \multicolumn{1}{l}{ 37.11} & \multicolumn{1}{l|}{ 79.48} & \multicolumn{1}{c|}{778.7M} & \multicolumn{1}{c|}{32} & 1h54m $\times$ 1 \\
\multicolumn{1}{c|}{2016} & \multicolumn{1}{c|}{DNN} & — & \multicolumn{1}{c|}{—} & 37.22 & \multicolumn{1}{l|}{ 79.28} &\multicolumn{1}{c|}{63.4M} & \multicolumn{1}{c|}{44} & \multicolumn{1}{l}{ 37.05} & \multicolumn{1}{l|}{ 79.59} &\multicolumn{1}{c|}{338.9M} & \multicolumn{1}{c|}{252} & 12m $\times$ 1 \\
\multicolumn{1}{c|}{2016} & \multicolumn{1}{c|}{WideDeep} & 37.44 & \multicolumn{1}{c|}{77.49} & 37.20 & \multicolumn{1}{l|}{ 79.29} & \multicolumn{1}{c|}{76.5M} & \multicolumn{1}{c|}{27} & \multicolumn{1}{l}{ 37.03} & \multicolumn{1}{l|}{ 79.57} & \multicolumn{1}{c|}{345.9M} & \multicolumn{1}{c|}{52} & 14m $\times$ 1 \\
\multicolumn{1}{c|}{2016} & \multicolumn{1}{c|}{IPNN} & 37.37 & \multicolumn{1}{c|}{78.68} & \textbf{37.12(4)} & \multicolumn{1}{c|}{\textbf{79.44(3)}} & \multicolumn{1}{c|}{62.7M} & \multicolumn{1}{c|}{18} & \multicolumn{1}{c}{\textbf{36.86(4)}} & \multicolumn{1}{c|}{\textbf{79.88(4)}} & \multicolumn{1}{c|}{336.1M} & \multicolumn{1}{c|}{36} & 11m $\times$ 1 \\
\multicolumn{1}{c|}{2016} & \multicolumn{1}{c|}{DeepCross} & 38.89 & \multicolumn{1}{c|}{76.43} & 37.21 & \multicolumn{1}{l|}{79.30} & \multicolumn{1}{c|}{63.1M} & \multicolumn{1}{c|}{36} & \multicolumn{1}{l}{ 37.00} & \multicolumn{1}{l|}{ 79.62} & \multicolumn{1}{c|}{342.6M} & \multicolumn{1}{c|}{45} & 12m $\times$ 1 \\
\multicolumn{1}{c|}{2017} & \multicolumn{1}{c|}{AFM} & 37.66 & \multicolumn{1}{c|}{77.40} &37.93 & \multicolumn{1}{l|}{ 78.23} & \multicolumn{1}{c|}{63.8M} & \multicolumn{1}{c|}{29} & \multicolumn{1}{l}{ 37.81} & \multicolumn{1}{l|}{ 78.40} & \multicolumn{1}{c|}{343.3M} & \multicolumn{1}{c|}{67} & 20m $\times$ 2 \\
\multicolumn{1}{c|}{2017} & \multicolumn{1}{c|}{NFM} & 37.61 & \multicolumn{1}{c|}{77.08} & 37.43 & \multicolumn{1}{l|}{ 78.94} & \multicolumn{1}{c|}{66.8M} & \multicolumn{1}{c|}{48} & \multicolumn{1}{l}{ 37.15} & \multicolumn{1}{l|}{ 79.40} & \multicolumn{1}{c|}{346.4M} & \multicolumn{1}{c|}{54} & 16m $\times$ 1 \\
\multicolumn{1}{c|}{2017} & \multicolumn{1}{c|}{DeepFM} & 37.42 & \multicolumn{1}{c|}{78.36} & 37.19 & \multicolumn{1}{l|}{ 79.30} & \multicolumn{1}{c|}{76.5M} & \multicolumn{1}{c|}{24} & \multicolumn{1}{l}{ 37.02} & \multicolumn{1}{l|}{ 79.62} & \multicolumn{1}{c|}{373.2M} & \multicolumn{1}{c|}{234} & 16m $\times$ 1 \\
\multicolumn{1}{c|}{2017} & \multicolumn{1}{c|}{DCN} & 37.21 & \multicolumn{1}{c|}{76.81} & 37.19 & \multicolumn{1}{l|}{ 79.31} & \multicolumn{1}{c|}{68.8M} & \multicolumn{1}{c|}{139} & \multicolumn{1}{l}{ 36.99} & \multicolumn{1}{l|}{ 79.65} & \multicolumn{1}{c|}{336.9M} & \multicolumn{1}{c|}{496} & 7m $\times$ 1 \\
\multicolumn{1}{c|}{2018} & \multicolumn{1}{c|}{FwFM} & 39.88 & \multicolumn{1}{c|}{74.06} & 37.44 & \multicolumn{1}{l|}{ 79.07} & \multicolumn{1}{c|}{60.0M} & \multicolumn{1}{c|}{15} & \multicolumn{1}{l}{ 37.24} & \multicolumn{1}{l|}{ 79.25} & \multicolumn{1}{c|}{334.9M} & \multicolumn{1}{c|}{15} & 11m $\times$ 1 \\
\multicolumn{1}{c|}{2018} & \multicolumn{1}{c|}{xDeepFM} & 37.37 & \multicolumn{1}{c|}{78.55} & 37.18 & \multicolumn{1}{l|}{ 79.33} & \multicolumn{1}{c|}{64.6M} & \multicolumn{1}{c|}{180} & \multicolumn{1}{l}{ 36.97} & \multicolumn{1}{l|}{ 79.67} & \multicolumn{1}{c|}{344.1M} & \multicolumn{1}{c|}{288} & 31m $\times$ 1 \\
\multicolumn{1}{c|}{2019} & \multicolumn{1}{c|}{HFM+} & — & \multicolumn{1}{c|}{—} & \textbf{37.14(5)} & \multicolumn{1}{c|}{\textbf{79.44(3)}} & \multicolumn{1}{c|}{66.2M} & \multicolumn{1}{c|}{39} & \multicolumn{1}{c}{\textbf{36.83(3)}} & \multicolumn{1}{c|}{\textbf{79.92(3)}} & \multicolumn{1}{c|}{355.3M} & \multicolumn{1}{c|}{92} & 43m $\times$ 1 \\
\multicolumn{1}{c|}{2019} & \multicolumn{1}{c|}{FGCNN} & 37.46 & \multicolumn{1}{c|}{78.83} & \textbf{37.11(3)} & \multicolumn{1}{c|}{\textbf{79.44(3)}} & \multicolumn{1}{c|}{146.7M} & \multicolumn{1}{c|}{52} & \multicolumn{1}{c}{\textbf{36.96(5)}} & \multicolumn{1}{c|}{\textbf{79.71(5)}} & \multicolumn{1}{c|}{374.5M} & \multicolumn{1}{c|}{84} & 2h10m $\times$ 1 \\
\multicolumn{1}{c|}{2019} & \multicolumn{1}{c|}{AutoInt+} & 38.11 & \multicolumn{1}{c|}{77.74} & 37.46 & \multicolumn{1}{l|}{ 79.02} & \multicolumn{1}{c|}{66.4M} & \multicolumn{1}{c|}{114} & \multicolumn{1}{l}{ 37.09} & \multicolumn{1}{l|}{ 79.53} & \multicolumn{1}{c|}{337.9M} & \multicolumn{1}{c|}{75} & 15m $\times$ 1 \\
\multicolumn{1}{c|}{2019} & \multicolumn{1}{c|}{FiGNN} & 38.25 & \multicolumn{1}{c|}{{{77.62}}} & 37.36 & \multicolumn{1}{l|}{ 79.15} & \multicolumn{1}{c|}{60.1M} & \multicolumn{1}{c|}{69} & \multicolumn{1}{l}{ 37.11} & \multicolumn{1}{l|}{ 79.44} & \multicolumn{1}{c|}{335.0M} & \multicolumn{1}{c|}{64} & 2h25m $\times$ 1 \\
\multicolumn{1}{c|}{2019} & \multicolumn{1}{c|}{ONN} & 39.45 & \multicolumn{1}{c|}{75.13} & \textbf{36.83(1)} & \multicolumn{1}{c|}{\textbf{79.92(1)}} & \multicolumn{1}{c|}{723.7M} & \multicolumn{1}{c|}{18} & \multicolumn{1}{c}{\textbf{36.77(2)}} & \multicolumn{1}{c|}{\textbf{80.01(2)}} & \multicolumn{1}{c|}{406.3M} & \multicolumn{1}{c|}{120} & 1h52 $\times$ 1 \\
\multicolumn{1}{c|}{2019} & \multicolumn{1}{c|}{FiBiNET} & 37.86 & \multicolumn{1}{c|}{78.32} & \textbf{37.05(2)} & \multicolumn{1}{c|}{\textbf{79.53(2)}} & \multicolumn{1}{c|}{89.5M} & \multicolumn{1}{c|}{72} & \multicolumn{1}{c}{\textbf{36.75(1)}} & \multicolumn{1}{c|}{\textbf{80.03(1)}} & \multicolumn{1}{c|}{395.9M} & \multicolumn{1}{c|}{54} & 36m  $\times$ 1 \\
\multicolumn{1}{c|}{2020} & \multicolumn{1}{c|}{LorentzFM} & 38.28 & \multicolumn{1}{c|}{77.75} & 37.56 & \multicolumn{1}{l|}{ 78.85} & \multicolumn{1}{c|}{60.0M} & \multicolumn{1}{c|}{5} & \multicolumn{1}{l}{ 37.42} & \multicolumn{1}{l|}{ 79.12} & \multicolumn{1}{c|}{334.9M} & \multicolumn{1}{c|}{9} & 46m $\times$ 17 \\
\multicolumn{1}{c|}{2020} & \multicolumn{1}{c|}{AFN+} & {37.18} & \multicolumn{1}{c|}{75.55} & 37.26 & \multicolumn{1}{l|}{ 79.29} & \multicolumn{1}{c|}{141.7M} & \multicolumn{1}{c|}{20} & \multicolumn{1}{l}{ 37.00} & \multicolumn{1}{l|}{ 79.65} & \multicolumn{1}{c|}{363.5M} & \multicolumn{1}{c|}{190} & 35m $\times$  1 \\
\multicolumn{1}{l|}{2020} & \multicolumn{1}{l|}{InterHAt} & 39.10 & \multicolumn{1}{c|}{75.82} & 37.49 & \multicolumn{1}{l|}{ 78.82} &  \multicolumn{1}{c|}{60.1M} & \multicolumn{1}{c|}{66} & \multicolumn{1}{l}{ 37.22} & \multicolumn{1}{l|}{ 79.27} & \multicolumn{1}{c|}{335.1M} & \multicolumn{1}{c|}{56} & 17m x 1 \\ \hline
\end{tabular}
% \vspace{-3ex}
\end{table*}
%%%%%%%%%%%%%%%%%%%%%%%%%%%%%%%%%%%%%%%%%%%%%%%%%%%%%%%%%%

%%%%%%%%%%%%%%%%%%%%%%%%%%%%%%%%%%%%%%%%%%%%%%%%%%%%
\begin{figure*}[!t]
\centering    %居中
\subfigure[Criteo dataset] %第一张子图
{
	\centering          %子图居中
	\includegraphics[width=0.488\textwidth]{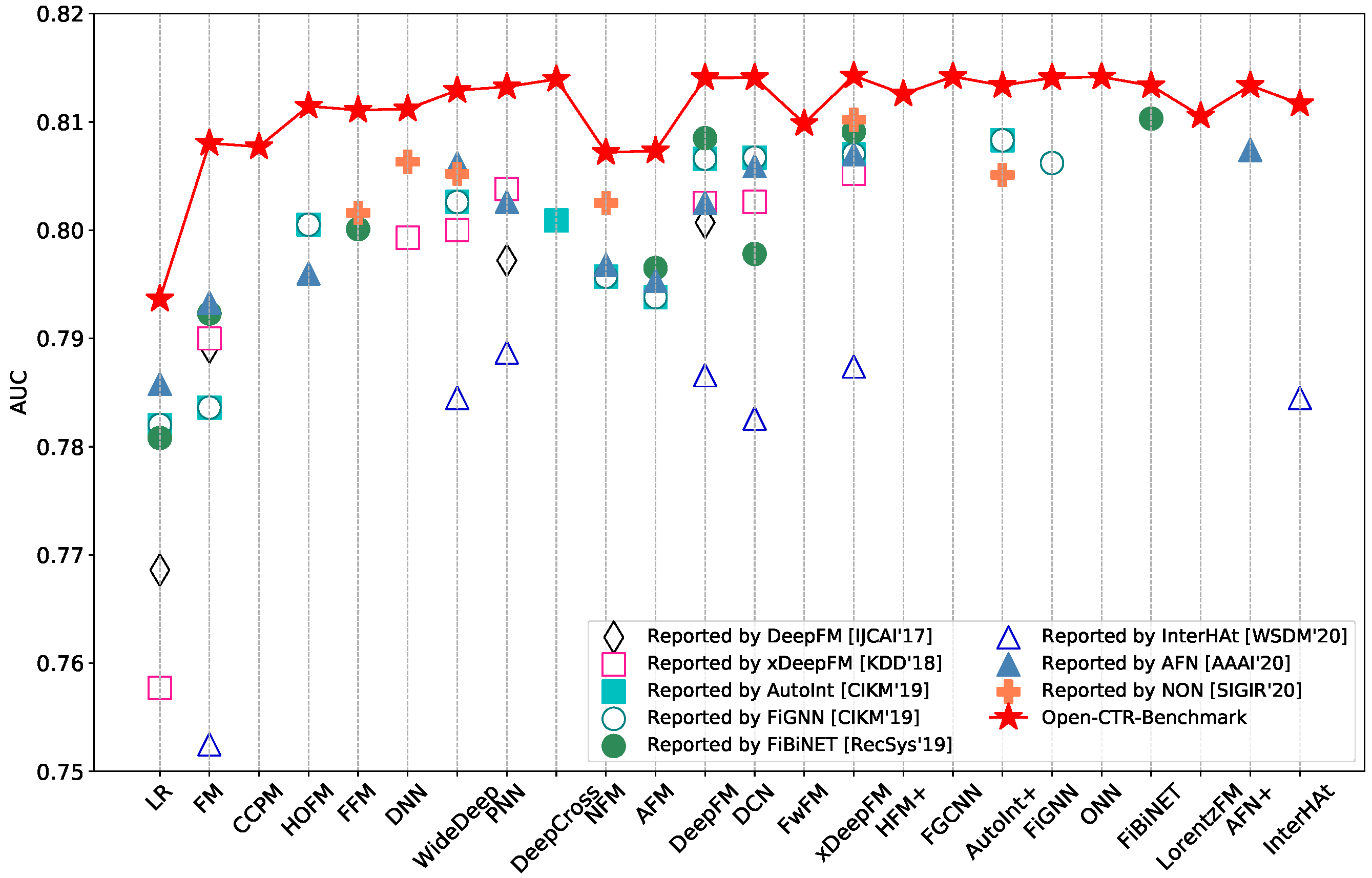}  
}
\subfigure[Avazu dataset] %第一张子图
{
	\centering      
	\includegraphics[width=0.488\textwidth]{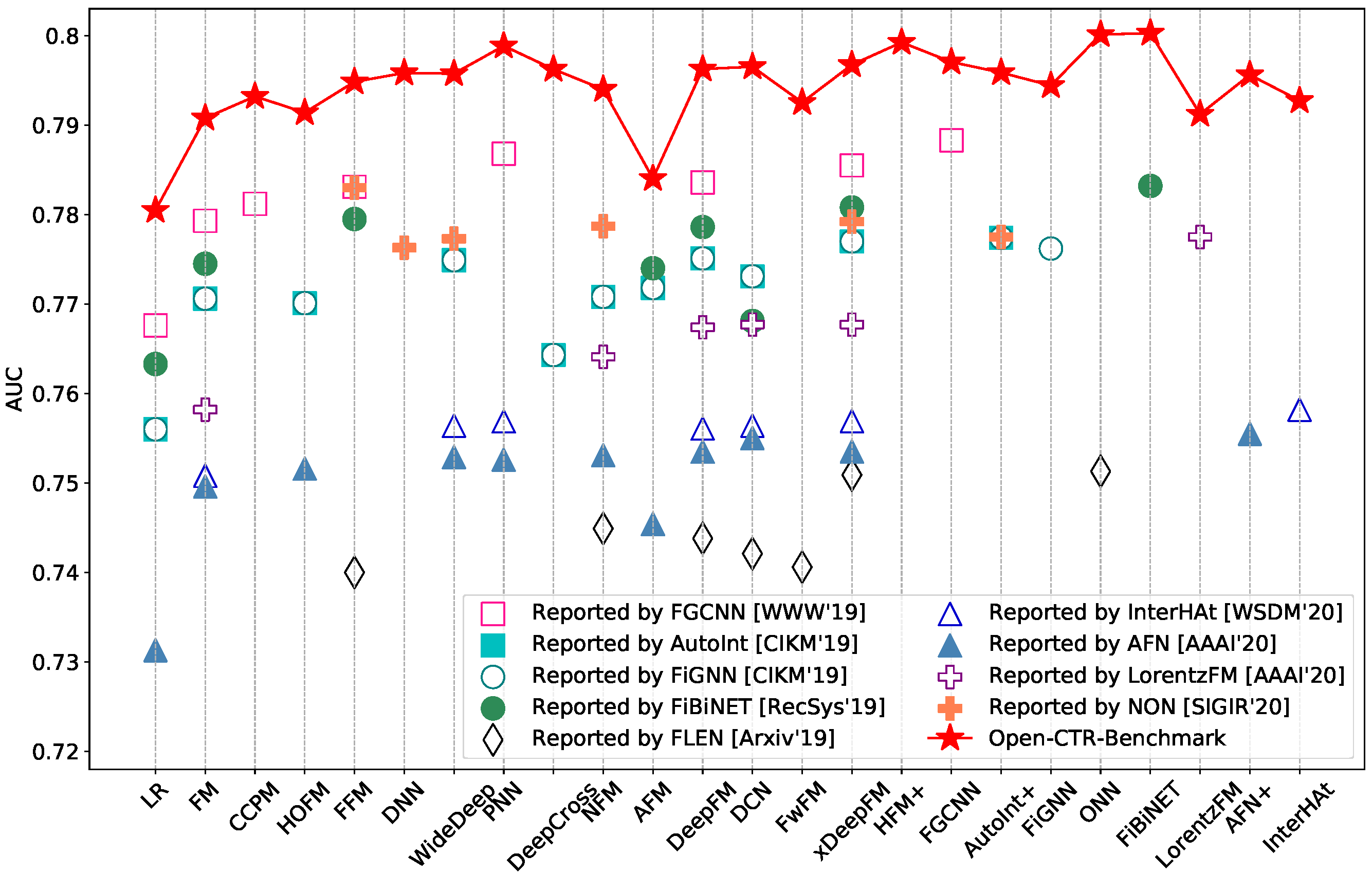}
}
\vspace{-1ex}
\caption{Comparison between our benchmarking results and those reported by the other papers}\label{fig:comparison}
\label{fig:compare} \vspace{-0.5ex}
\end{figure*}
%%%%%%%%%%%%%%%%%%%%%%%%%%%%%%%%%%%%%%%%%%%%%%%%%%%%
\begin{table}[!t]
\centering
\caption{Results before and after retuning hyper-parameters (Logloss $\times10^{-2}$).}\label{tab:retuning}
\begin{tabular}{c|c|cc|cc}
\hline
\multirow{2}{*}{Model} & \multirow{2}{*}{Setting} & \multicolumn{2}{c|}{Criteo\_x4} & \multicolumn{2}{c}{Avazu\_x4} \\ \cline{3-6} 
 &  & Logloss & AUC(\%) & Logloss & AUC(\%) \\ \hline
\multirow{3}{*}{AutoInt+} & Reported & 44.34 & 80.83 & 38.11 & 77.74 \\
 & Rerun & 44.25 & 80.92 & 37.32 & 79.13 \\
 & Retuned & \textbf{43.85} & \textbf{81.34} & \textbf{37.09} & \textbf{79.53} \\ \hline
\multirow{3}{*}{FiGNN} & Reported & 44.53 & 80.62 & 38.25 & 77.62 \\
 & Rerun & 44.28 & 80.89 & 37.44 & 78.93 \\
 & Retuned & \textbf{43.79} & \textbf{81.41} & \textbf{37.11} & \textbf{79.44} \\ \hline
\multirow{3}{*}{FiBiNET} & Reported & 44.23 & 81.03 & 37.86 & 78.32 \\
 & Rerun & 44.31 & 80.88 & 37.24 & 79.37 \\
 & Retuned & \textbf{43.86} & \textbf{81.34} & \textbf{36.75} & \textbf{80.03} \\ \hline
\multirow{3}{*}{AFN+} & Reported & 0.4451 & 80.74 & 0.3718 & 75.55 \\
 & Rerun & 44.32 & 80.84 & 37.15 & 79.39 \\
 & Retuned & \textbf{43.87} & \textbf{81.34} & \textbf{37.00} & \textbf{79.65} \\ \hline
\end{tabular}
\end{table}
%%%%%%%%%%%%%%%%%%%%%%%%%%%%%%%

\subsection{Results Analysis}\label{sec:RQ1}
In this section, we report our benchmarking results of 24 models, as shown in Table~\ref{tab:benchmarking}. To be specific, the "Best Reported" column shows the best results that we select from those reported by existing studies on Criteo and Avazu datasets. We report our benchmarking results w.r.t. logloss and AUC on four dataset settings, i.e., Criteo
\_x4\_001, Criteo\_x4\_002, Avazu\_x4\_001, and Avazu\_x4\_002. For model efficiency, we also report training time, in terms of time per epoch and number of epochs, for Criteo\_x4\_002 and Avazu\_x4\_002. In addition, "\#Params" denotes the number of parameters used in each model. "\#Runs" records the number of experiments we run with grid search for model tuning. Note that the "\#Runs" values typically depend on the number of hyper-parameters to tune in a model. The large number of runs ($\sim$73 on average) reveals that the models have been well tuned in our benchmarking. Moreover, we run the experiments on Criteo\_x4\_002 and Avazu\_x4\_002 first, so we perform a smaller number of experiments to tune models on Criteo\_x4\_001 and Avazu\_x4\_001. From the benchmarking results in Table~\ref{tab:benchmarking}, we have the following surprising observations:

\begin{itemize}[leftmargin=*]
\item The best results reported by existing studies show certain inconsistency. For example, InterHAt has worse performance than LR on both datasets; DeepCross performs worse than LR on Avazu as well. They are largely due to the fact that different data splitting or preprocessing steps are usually applied even on the same datasets. This reveals that standardized data splitting and preprocessing is necessary to make the results directly comparable among models.
\item After model retuning on our dataset settings, we mostly obtain better performance than the best reported results. While their data settings may be different from what we described in Section~\ref{sec:protocol}, these results could serve as a good reference for our study. Considering that we follow the same data splitting as~\cite{autoint}, the large improvements indicate that our modified data preprocessing is indeed more appropriate for model evaluation, resulting in stronger baselines results to build. This is also more practical, conforming to the real case where data preprocessing is usually heavily tuned. 
\item Our benchmarking follows the same evaluation protocol to make the results comparable. Yet, after exhaustive retuning, we found that the differences among the state-of-the-art models become small. For example, IPNN, DeepFM, DCN, xDeepFM, and ONN all achieve the same level of accuracy ($\sim$0.814 AUC) on Criteo, while DNN, DeepFM, DCN and xDeepFM attain comparable performance on Avazu. We run many experiments with different hyper-parameters, but do not obtain sufficiently distinct results. Especially, we run DCN for 544 and 496 experiments on Criteo\_x4\_002 and Avazu\_x4\_002 respectively, yet only make indistinguishable differences compared to DeepFM. The same happens for xDeepFM. Moreover, some recent models, such as InterHAt, AFN+, and LorentzFM obtain even worse results than some previous state-of-the-arts. Our results highlight the value of open benchmarking, which aims to provide reproducible and strong baseline results for future research. 
\item We also make a clear comparison between our benchmarking results and those reported by the other papers in Figure~\ref{fig:comparison}. For each dataset, we plot the AUC results from 8$\sim$9 existing papers. We can see that the results vary a lot among different papers due to unknown data splits and preprocessing. The X-axis is arranged sequentially according to the publication year, but there is no obvious pattern on model performance improvement. Some recent models only obtain diminished improvements and sometimes even lead to performance drops. Remarkably, our benchmarking presents the best results reported so far for all models.
\item Memory consumption and model efficiency are another two important aspects to industrial CTR prediction tasks. As shown in the table, some models run very slowly (hours per epoch) due to the use of convolution networks (e.g., CCPM, FGCNN, HFM+), fieldwise interactions (e.g., FFM, ONN), graph neural networks (e.g., FiGNN), and so on. Some others have more parameters, such as FFM, ONN, FGCNN, etc. These drawbacks might hinder their practical use in industry. 
%Note that these numbers provide a reference only, because they depends heavily on the model hyper-parameters. For example, it is possible to reduce the number of parameters and training time at the expense of decreased accuracy, by reducing the number of layers or setting a smaller hidden size.  
% \item AutoInt, 
% \item FiGNN
% \item NFM and AFM
\end{itemize}

\subsubsection{Model retuning}
To further demonstrate the necessity to retune baseline models, in Table~\ref{tab:retuning}, we present the results of four representative models in three settings: 1) The \textit{Reported} setting denotes the results reported by the corresponding papers. 2) The \textit{Rerun} setting indicates the cases that we rerun the experiments on our data splits according to the hyper-parameters given in the original papers. 3) The \textit{Retuned} setting shows the results achieved after extensive hyper-parameter tuning. We can see that even on the same datasets, directly reusing the original hyper-parameters to rerun the experiments could bring large performance gaps in new data settings (i.e., Criteo\_x4\_002 and Avazu\_x4\_002 in our case). After model retuning, we achieve quite large improvements (up to 5\textperthousand) over the original hyper-parameters, considering that 1\textperthousand ~improvement is deemed to be significantly practical. This indicates that it is necessary to retune the hyper-parameters when testing a model on a new data split (even for the same dataset). However, it is not uncommon to find some studies (e.g., \cite{FiBiNET}) that choose to follow the baseline hyper-parameters used in the papers for fairness of comparison, but they experiment on a different data split. A common benchmark is thus desired to alleviate this issue.

\subsection{Key Factors to Performance Tuning}\label{sec:performance_tuning}
During our benchmarking work, we also identify some key factors that are critical for performance tuning. 
\begin{itemize}[leftmargin=*]
\item\textbf{Data preprocessing}. Data often determine the upper bound of a model. However, existing work seldom tunes the min\_counts thresholds for category features during data preprocessing. In our work, we set an appropriate threshold for infrequent feature filtering, which yields much better performance. 
\item\textbf{Batch size}. We observe that a large batch size usually results in faster training and better performance. For example, we set it to 10000 if the GPU does no raise an OOM error. 
\item\textbf{Embedding size}. While existing work usually set it to 10 or 16 in the experiments, we also experiment with other settings by using a larger embedding size (e.g., 40) within the GPU memory constraints. 
\item\textbf{Regularization weight and dropout rate}. Regularization and dropout are two key techniques to reduce model overfitting. They have a large impact on the performance of CTR prediction models. We exhaustively search the optimal value within a range. 
\item\textbf{Batch normalization}. In some cases, adding batch normalization between hidden layers of a DNN model can further boost prediction performance. 
\end{itemize}
%For detailed benchmarking settings, please refer to the source code and configuration files on our \texttt{Open-CTR-Benchmark} website.

%% file: sections/5_relatedwork.tex
\section{Discussion}\label{sec:discussion}
In this section, we discuss the limitations and potential directions for further exploration. 

\textbf{More datasets}: In this work, we evaluate and benchmark existing CTR prediction models on two widely-used datasets, Criteo and Avazu. However, both datasets are anonymized and there is a lack of explicit user field and item field information. Therefore, they are able to be used to benchmark some models that require explicit user-item interaction (e.g., FLEN~\cite{FLEN}) and user behavior sequence information (e.g., DIN~\cite{DIN}). We plan to extend more datasets from industrial-scale applications to make it a more comprehensive open benchmark for CTR prediction. %We emphasize that although some studies employ small datasets (e.g., Frappe) for CTR prediction evaluation, we advocate for more research on industrial-scale datasets. 

%we disagree with them because: 1) these datasets are small and usually share different characteristics with real industrial settings (e.g., billions of samples with millions of features~\cite{WideDeep}); 2) Due to the lack of real negative samples in Frappe, random negative sampling is applied for both training and testing. However, this may lead to a totally different sample distribution with CTR prediction tasks, since most irrelevant candidate items are filtered during the matching phase. Therefore, 

%2) Movielens has a long time span (over years) and very few feature fields (e.g., only 3 fields used in~\cite{AFM}), which is not representative of most real-time recommender systems; 

\textbf{Data splits}: To keep it consistent with most existing research, we split the datasets randomly to benchmark CTR prediction models. We do so with the following consideration. With randomly splitting, the data distributions among train, validation, and test sets are more consistent. This helps to better reveal the effectiveness of a CTR prediction model on capturing feature interactions, because in production it is necessary to perform CTR calibration after prediction if the train-test distributions vary largely. As part of future work, we will evaluate the models by splitting data sequentially over time and also perform CTR calibration as necessary.

%\textbf{User interests modeling}: Recently, some work~\cite{DIN, DIEN, DSIN} proposes to model user history interaction sequence to enhance CTR prediction. However, we cannot evaluate these models on Criteo and Avazu datasets, due to the lack of user id and item id information in the two mostly-widely used datasets. We will extend more datasets in FuxiCTR to benchmark these models.

\textbf{Efficiency benchmarking}: Current models for CTR prediction has become more and more complex in structure, after using components such as attention~\cite{autoint, InterHAt}, convolution~\cite{CNN-FeatureGen}, and graph neural network~\cite{FiGNN}. In this version, we mainly evaluate the efficiency of these models through their training time. It is also desirable to test their inference time in future. Due to the strict latency constraints of CTR prediction in real-time applications, efficiency benchmarking would not only help practitioners choose an appropriate model, but also facilitate researchers to design effective yet efficient models. 

% \textbf{Feature construction and selection }: 

\textbf{Auto-tuning of hyper-parameters}: As the experimental results shown in Section~\ref{sec:exp}, hyper-parameter tuning is critical to the performance of CTR prediction models. How to quickly find the optimal hyper-parameter for a given model remains an open research problem. When data evolve with time, model hyper-parameters re-tuning is also required to adapt to the new data distribution. In our benchmark, we mainly apply grid search to find the best hyper-parameters of each model. It is highly expected to explore some advanced AutoML techniques (e.g., bayesian optimization in NNI~\cite{NNI}) to further boost hyper-parameter tuning process in future.

% \textbf{FuxiCTR versions}: 

% \textbf{Multimodal}: 

% \textbf{Embedding}: 

% \subsection{Debiasing}: 

% \subsection{Multi-Task Learning}

\section{Related Work}\label{sec:relatedwork}
% In this section, we review the related work on CTR prediction and open benchmarking.

\subsection{CTR Prediction}
During the last decade, CTR prediction models have been widely studied and evolved through several generations from linear models~\cite{FTRL}, to factorization machines~\cite{FM}, and to deep learning-based models~\cite{DNN}. 
We have introduced some representative models in Section~\ref{sec_background}. Here, we present a review of more related studies on CTR prediction, which are summarized into the following categories. %The goal of this work is to provide a standardized benchmarking for CTR prediction and guide future research in this area. 

\textbf{Feature interaction learning}. 
While simple linear models such as LR~\cite{LR} and FTRL~\cite{FTRL} have been widely used due to their simplicity and efficiency, they have difficulty capturing non-linear feature mappings and conjunctions. He et al.~\cite{GBDTLR} propose the GBDT + LR approach that applies Gradient Boosting Decision Tree (GBDT) to extract meaningful feature conjunctions. 

FM~\cite{FM} is an effective model that captures pairwise feature interactions via inner products of feature vectors. Due to its success, many follow-up models have been proposed from different aspects, such as field awareness (e.g., FFM~\cite{FFM}, FwFM~\cite{FwFM}), importance of feature interactions (e.g., AFM~\cite{AFM}, IFM~\cite{IFM}), outer-products based interaction (HFM~\cite{HFM}), robustness (RFM~\cite{RFM}), and interpretability (SEFM~\cite{SEFM}). However, it is non-trivial for these models to capture high-order feature interactions in practice~\cite{HighFM}.

Recently, deep learning has become a popular technique in recommender systems~\cite{DeepSurvey}, yielding an abundance of deep models for CTR prediction, including YoutubeDNN~\cite{DNN}, Wide\&Deep~\cite{WideDeep}, PNN~\cite{PNN}, DeepFM~\cite{DeepFM}, DistillCTR~\cite{DistillCTR}, etc. Some of them aim to capture different orders of feature interactions explicitly (e.g., DCN~\cite{DCN}, xDeepFM~\cite{xDeepFM}). Some other models explore the use of convolutional networks (e.g., CCPM~\cite{CCPM}, FGCNN~\cite{CNN-FeatureGen}), recurrent networks (e.g., ~\cite{DSIN,DIEN}), or attention networks (e.g., AutoInt~\cite{autoint}, FiBiNET~\cite{FiBiNET}) to learn implicit feature interactions.   

\textbf{Behaviour sequence modeling}. 
The history behaviours of users have a large effect on predicting the click probability on the next item. To better capture such history behaviours (e.g., item purchase sequences), some recent studies propose user interests modeling for CTR prediction via attention, LSTM, GRU, and memory networks. Typical examples include DIN~\cite{DIN}, DIEN~\cite{DIEN}, DSIN~\cite{DSIN}, HPMN~\cite{HPMN}) and DSTN~\cite{DSTN}.

\textbf{Multi-task learning}. 
In many recommender systems, users may have diverse behaviors beyond clicks, such as browsing, favorite, add-to-cart, and purchase. To improve the performance of CTR prediction, it is desirable to leverage other types of user feedbacks to enrich the supervision signals for CTR prediction. Towards this goal, some work proposes multi-task learning models to learn task relationships among different user behaviors, such as ESMM~\cite{ESMM}, MMoE~\cite{MMoE} and PLE~\cite{PLE}.

\textbf{Multi-modal learning}. 
Nowadays, multi-modal information contents are prevalent in recommender systems, such as e-commerce, news feeds, micro-blogs, and micro-videos. How to employ the rich multi-modal information (e.g., text, image, and video) of items to enhance CTR prediction models is an important research problem that needs more exploration. Some pioneer work (e.g., ~\cite{UNBERT,WWW_multimodal_ctr,Telepath,ImageMatters,IMRec}) demonstrates the effectiveness to incorporate multi-modal content features in CTR prediction.

% \subsubsection{Model Ensemble and Distillation}
% Although model ensemble is a powerful technique to boost prediction accuracy, the high complexity in such ensembles (e.g., over 100 models ensembled in the Netflix prize~\cite{NetflixPrize}) hinders the adoption in the industry. To restrict the model complexity for online deployment, most industrial studies reported by Facebook~\cite{GBDTLR}, Google~\cite{GoogleEnsemble}, and Microsoft~\cite{EnsembleBing,DeepGBM} focus on the ensemble of only two models. Especially, some deep models mentioned above (e.g., DCN~\cite{DCN}) can be viewed as a type of ensemble of two models via joint training.

% \subsubsection{Debiasing in CTR Prediction}

% PAL, CIKM
% WWW_debias
% A General Knowledge Distillation Framework for Counterfactual Recommendation via Uniform Data
% Sampling-bias-corrected neural modeling for large corpus item recommendations

% Transfer Learning
% Online/Lifelong Learning
% Cold-start
% Embedding

\subsection{Benchmarking and Reproducibility}
With the prevalence of deep learning, new models are emerging at an increasingly rapid pace. There is a high demand for an open benchmark to fairly compare against baseline models. Open benchmarking is valuable to promote research progress. For example, ImageNet~\cite{ImageNet} and GLUE~\cite{GLUE} are two well-known benchmarks that contribute much to the progress in computer vision and natural language processing, respectively. In recommender systems, some datasets (e.g., Criteo and Avazu) are widely used. However, there is still a lack of standardized evaluation protocols, which results in the inconsistency and non-reproducibility issues of existing studies~\cite{Recsys_worryanalysis}. Notably, a concurrent work~\cite{RecSysBench} also reports the benchmarking results on some classic recommendation models. Yet, their work fails to provide detailed configurations and hyper-parameter settings to allow reproducibility. In this work, we take an important step towards reproducible research by building the first open benchmark for CTR prediction as well as releasing the benchmarking results for over 20 models. More importantly, we provide all the intermediate artifacts (e.g., reproducing steps, running logs) to ensure reproducibility of our results.

%% file: sections/6_conclusion.tex
\section{Conclusion}\label{sec:conclusion}
In this paper, we present FuxiCTR-Bench as the first open benchmark for CTR prediction. We aim to alleviate the issues of non-reproducible and inconsistent results raised in current studies. We standardize the evaluation protocols and evaluate 24 existing models by running over 7,000 experiments for more than 12,000 GPU hours on two widely-used real-world datasets. We provide the most comprehensive benchmarking results that compare existing models in a rigorous manner. The results show that the difference between many models is smaller than expected and that inconsistent results exist in existing papers. We believe that our benchmark would not only drive more reproducible research but also help new beginners to learn the state-of-the-art CTR prediction models. 
%The results show that the difference between many models are smaller than expected and the results are even inconsistent with what reported in the original paper. Our benchmark is a valuable work that providing a methodology for researchers to verify the effectiveness of new CTR models.